\documentclass[english]{article}
\usepackage{times}
\usepackage[T1]{fontenc}
\usepackage[latin1]{inputenc}
\usepackage{geometry}
\usepackage{color}
\usepackage{graphicx}
\usepackage{setspace}

\makeatletter

\providecommand{\tabularnewline}{\\}

\usepackage{axodraw}

\usepackage{babel}
\makeatother
\begin{document}

\title{Prospects to search for E6 isosinglet quarks in ATLAS}

\maketitle
\begin{center}\textcolor{black}{\large R. Mehdiyev}%
\footnote{\textcolor{black}{Universit{\'e} de Montr{\'e}al, D{\'e}partement
de Physique, Montr{\'e}al, Canada. }%
}\textcolor{black}{$^{,}$}%
\footnote{\textcolor{black}{Institute of Physics, Academy of Sciences, Baku,
Azerbaijan.}%
}\textcolor{black}{\large , S. Sultansoy}\textcolor{black}{$^{2,}$}%
\footnote{\textcolor{black}{Gazi University, Physics Department, Ankara, Turkey.}%
}\textcolor{black}{\large , G. Unel}\textcolor{black}{}%
\footnote{\textcolor{black}{CERN, Physics Department, Geneva, Switzerland.}%
}\textcolor{black}{$^{,}$}%
\footnote{\textcolor{black}{University of California at Irvine, Physics Department,
USA. }%
}\textcolor{black}{\large , M. Yilmaz}\textcolor{black}{$^{3}$}\end{center}

\begin{abstract}
We consider pair production of new down-type isosinglet quarks originating
from $E_{6}$, which is the favorite gauge symmetry group in superstring
inspired GUT models. The study concentrates on the possibility of
observing the pair production of the lightest of the new quarks, $D$,
in the ATLAS detector at the forthcoming LHC accelerator, in the channel
$D\bar{D}\rightarrow ZjZj$ . Both signal and background events are
studied using tree level event generators based on Monte Carlo techniques.
The detector effects are taken into account using the ATLAS fast simulation
tool, ATLFAST. It is shown that ATLAS can observe the $D$ quark within
the first year of low luminosity LHC operation if its mass is less
than 650~GeV. For the case of two years of full luminosity running,
1~TeV can be reached with about three sigma significance. 
\end{abstract}

\section{Introduction}

If observed, the long awaited discovery of the Higgs particle at the
LHC experiments \cite{R-atlas-tdr,R-CMS-tdr} will complete the validation
of the basic principles of the Standard Model (SM). However, the well
known deficiencies of the SM, such as the arbitrariness of the fermion
mass spectrum and mixings, the number of families, the real unification
of the fundamental interactions and the origin of baryon asymmetry
of the universe require extensions of the SM to achieve more complete
theories. In general, these extensions predict the existence of new
fundamental particles and interactions. The forthcoming LHC will give
the opportunity to explore new colored particles and their interactions
to test these predictions. Three types of new quarks (see \cite{R-Classification}
for a general classification) are of special interest: the fourth
SM family quarks, up type and down type weak isosinglet quarks. The
existence of the fourth SM family is favored by flavor democracy (see
\cite{R-democracy} and references therein for details), Q=2/3 quarks
are predicted by the little Higgs model \cite{LittleHiggs} and $Q=-1/3$
isosinglet quarks are predicted by Grand Unification Theories (GUTs),
with $E_{6}$ as the unification group \cite{R-e6}. The GUT models
permit solving at least two of the above mentioned problems, namely,
the complete unification of the fundamental interactions (except gravity)
and the baryon asymmetry of the observed universe by merging strong
and electroweak interactions in a single gauge group. Theories adding
gravity to the unification of fundamental forces, superstring and
supergravity theories \cite{R-String} also favor $E_{6}$ as a gauge
symmetry group when compactified from 10 (or 11) dimensions down to
the 3+1 that we observe (see \cite{R-hewet-rizzo} and references
therein for a review of $E_{6}$ GUTs).

For LHC, the production and observation of the first and second type
of new quarks have been investigated in \cite{R-atlas-tdr,R-4thfam}
and \cite{LittleHiggs,R-aguilar} respectively. In this work, we study
the possibility to observe the third, down type isosinglet quarks
predicted by the $E_{6}$-GUT model, at LHC in general, and, specifically
in the ATLAS experiment\cite{R-atlas-tdr} using the 4 lepton and
2 jet channel. The current experimental limit on the mass of an isosinglet
quark is $m>199$~GeV~\cite{PDG}. The detailed study for down type
isosinglet quark signatures at the Tevatron has been recently performed
in \cite{Rosner} and it has been shown that the upgraded Tevatron
would allow a mass reach up to about 300~GeV.

\section{The Model}

If the group structure of the SM, $SU_{C}(3)\times SU_{W}(2)\times U_{Y}(1)$,
originates from the breaking of the $E_{6}$ GUT scale down to the
electroweak scale, then the quark sector of the SM is extended in
the following manner:

\begin{equation}
\left(\begin{array}{c}
u_{L}\\
d_{L}\end{array}\right),u_{R},d_{R},D_{L},D_{R}\,;\quad\left(\begin{array}{c}
c_{L}\\
s_{L}\end{array}\right),c_{R},s_{R},S_{L},S_{R}\,;\quad\left(\begin{array}{c}
t_{L}\\
b_{L}\end{array}\right),t_{R},b_{R},B_{L},B_{R}\quad.\label{quarks}\end{equation}
As shown, each SM family is extended by the addition of an isosinglet
quark. The new quarks are denoted by letters $D$, $S$, and $B$.
The mixings between these and SM down type quarks is responsible for
the decays of the new quarks.

These mixings increase the number of angles ($N_{\Theta}$) and phases
($N_{\Phi}$) with respect to the SM CKM matrix. For a general multi-quark
model, one has \cite{N-params} :

\begin{eqnarray}
N_{\Theta} & = & N\times(l+m-\frac{3N+1}{2})\label{eq:num-params}\\
N_{\Phi} & = & (N-1)\times(l+m-\frac{3N+2}{2})\nonumber \end{eqnarray}

where $l$ and $m$ are the numbers of the up-type and down-type quarks
respectively; and $N$ is the number of $SU(2)_{W}$ doublets formed
by left handed quarks. In the case of the $E_{6}$ model, we have
$m=2l=2N=6$ and Eq. (\ref{eq:num-params}) yields $N_{\Theta}=12$
and $N_{\Phi}=7$ . The special case $m=l+1=N+1=4$, considered in
\cite{Rosner}, yields $N_{\Theta}=6$ and $N_{\Phi}=3$ which coincides
with the number of parameters in the Little Higgs models with one
additional isosinglet up-type quark \cite{LittleHiggs} . 

In this study, the intrafamily mixings of the new quarks are assumed
to be dominant with respect to their inter-family mixings. In addition,
as for the SM hierarchy, the $D$ quark is taken to be the lightest
one. The usual CKM mixings, represented by superscript $\theta$,
are taken to be in the up sector for simplicity of calculation (which
does not affect the results). Therefore, the Lagrangian relevant for
the decay of the $D$ quark becomes \cite{R-e6-orhan-metin} :

\begin{eqnarray}
{\cal {L}_{D}} & = & \frac{\sqrt{4\pi\alpha_{em}}}{2\sqrt{2}\sin\theta_{W}}\left[\bar{u}^{\theta}\gamma_{\alpha}\left(1-\gamma_{5}\right)d\cos\phi+\bar{u}^{\theta}\gamma_{\alpha}\left(1-\gamma_{5}\right)D\sin\phi\right]W^{\alpha}\label{lagrangian}\\
 & - & \frac{\sqrt{4\pi\alpha_{em}}}{4\sin\theta_{W}}\left[\frac{\sin\phi\cos\phi}{\cos\theta_{W}}\bar{d}\gamma_{\alpha}\left(1-\gamma_{5}\right)D\right]Z^{\alpha}+h.c.\nonumber \end{eqnarray}

\begin{figure}

\caption{The width of $D$ quark as a function of its mass ($\sin\phi$ =
0.05 )\label{cap:The-width-of-D}}

\begin{center}\includegraphics[%
  scale=0.45]{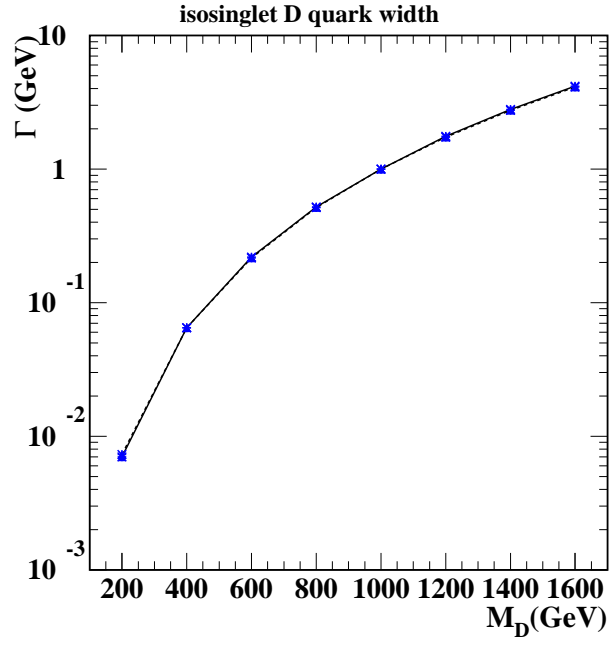}\end{center}
\end{figure}

The measured values of $V_{ud}$,$V_{us}$,$V_{ub}$ constrain the
$d$ and $D$ mixing angle $\phi$ to $|\sin\phi|$ $\leq0.07$~
assuming the squared sum of row elements of the new $3\times4$ CKM
matrix give unity (see\cite{PDG} and references therein for CKM matrix
related measurements). The total decay width and the contribution
by neutral and charged currents were already estimated in \cite{R-e6-orhan-metin}.
As reported in this work, the $D$ quark decays through a $W$ boson
with a branching ratio of 67\% and through a $Z$ boson with a branching
ratio of 33\%. The total width of the $D$ quark as a function of
its mass is shown in Fig.~\ref{cap:The-width-of-D} for the illustrative
value of $\sin\phi=0.05$. It is seen that the $D$~quark has a rather
narrow width and becomes even narrower with decreasing value of $\phi$
since it scales through a $\sin^{2}\phi$ dependence. If the Higgs
boson exists, in addition to these two modes, $D$ quark might also
decay via the $D\rightarrow H\, d$ channel which is available due
to $D-d$ mixing. The branching ratio of this channel for the case
of $m_{H}=120$~GeV and $\sin\phi=0.05$ is calculated to be about
25\%, reducing the branching ratios of the previously discussed neutral
and charged channels to 50\% and 25\%, respectively \cite{Rosner,e6-higgs}.
However, this study will not take into account the possible existence
of the Higgs boson and will concentrate on the pair production of
the $D$~quarks which is approximately independent of the value of
$\sin\phi$. One should note that the additional consideration of
single production of $D$$\;$quark increases the overall $D$$\;$quark
production rate; for example by about 40\% for $m_{D}=800$~GeV if
$\sin\phi=0.05$. However, final state particles and SM backgrounds
are different from the pair production case, making it a rather different
process to study.

\section{Pair Production at LHC - signal at generator level}

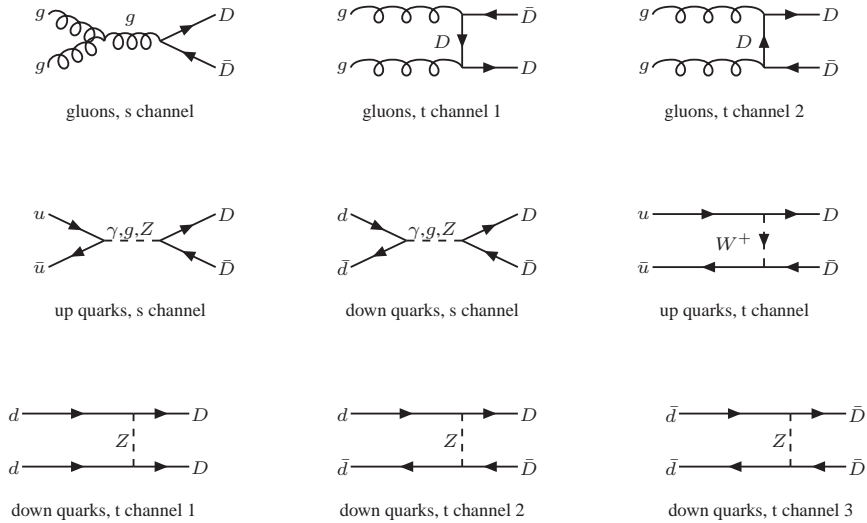
\begin{figure}

\caption{The tree level Feynman diagrams for the pair production of isosinglet
quarks \label{cap:signal-feynman}}

\begin{center}
{
\unitlength=1.0 pt
\SetScale{1.0}
\SetWidth{0.7}      
\scriptsize    
{} \qquad\allowbreak
\begin{picture}(95,79)(0,0)
\Text(15.0,70.0)[r]{$g$}
\Gluon(16.0,70.0)(37.0,60.0){3}{3}
\Text(15.0,50.0)[r]{$g$}
\Gluon(16.0,50.0)(37.0,60.0){3}{3}
\Text(47.0,66.0)[b]{$g$}
\Gluon(37.0,60.0)(58.0,60.0){3}{3}
\Text(80.0,70.0)[l]{$D$}
\ArrowLine(58.0,60.0)(79.0,70.0) 
\Text(80.0,50.0)[l]{$\bar{D}$}
\ArrowLine(79.0,50.0)(58.0,60.0) 
\Text(47,30)[b] {gluons, s channel}
\end{picture} \ 
{} \qquad\allowbreak
\begin{picture}(95,79)(0,0)
\Text(15.0,70.0)[r]{$g$}
\Gluon(16.0,70.0)(58.0,70.0){3}{3}
\Text(80.0,70.0)[l]{$\bar{D}$}
\ArrowLine(79.0,70.0)(58.0,70.0) 
\Text(54.0,60.0)[r]{$D$}
\ArrowLine(58.0,70.0)(58.0,50.0) 
\Text(15.0,50.0)[r]{$g$}
\Gluon(16.0,50.0)(58.0,50.0){3}{3}
\Text(80.0,50.0)[l]{$D$}
\ArrowLine(58.0,50.0)(79.0,50.0) 
\Text(47,30)[b] {gluons, t channel 1}
\end{picture} \ 
{} \qquad\allowbreak
\begin{picture}(95,79)(0,0)
\Text(15.0,70.0)[r]{$g$}
\Gluon(16.0,70.0)(58.0,70.0){3}{3}
\Text(80.0,70.0)[l]{$D$}
\ArrowLine(58.0,70.0)(79.0,70.0) 
\Text(54.0,60.0)[r]{$D$}
\ArrowLine(58.0,50.0)(58.0,70.0) 
\Text(15.0,50.0)[r]{$g$}
\Gluon(16.0,50.0)(58.0,50.0){3}{3}
\Text(80.0,50.0)[l]{$\bar{D}$}
\ArrowLine(79.0,50.0)(58.0,50.0) 
\Text(47,30)[b] {gluons, t channel 2}
\end{picture} \ 
}
\vskip-1cm
{
\unitlength=1.0 pt
\SetScale{1.0}
\SetWidth{0.7}      
\scriptsize    
{} \qquad\allowbreak
\begin{picture}(95,79)(0,0)
\Text(15.0,70.0)[r]{$u$}
\ArrowLine(16.0,70.0)(37.0,60.0) 
\Text(15.0,50.0)[r]{$\bar{u}$}
\ArrowLine(37.0,60.0)(16.0,50.0) 
\Text(47.0,61.0)[b]{$\gamma$,$g$,$Z$}
\DashLine(37.0,60.0)(58.0,60.0){3.0} 
\Text(80.0,70.0)[l]{$D$}
\ArrowLine(58.0,60.0)(79.0,70.0) 
\Text(80.0,50.0)[l]{$\bar{D}$}
\ArrowLine(79.0,50.0)(58.0,60.0) 
\Text(47,30)[b] {up quarks, s channel}
\end{picture} \ 
{} \qquad\allowbreak
\begin{picture}(95,79)(0,0)
\Text(15.0,70.0)[r]{$d$}
\ArrowLine(16.0,70.0)(37.0,60.0) 
\Text(15.0,50.0)[r]{$\bar{d}$}
\ArrowLine(37.0,60.0)(16.0,50.0) 
\Text(47.0,61.0)[b]{$\gamma$,$g$,$Z$}
\DashLine(37.0,60.0)(58.0,60.0){3.0} 
\Text(80.0,70.0)[l]{$D$}
\ArrowLine(58.0,60.0)(79.0,70.0) 
\Text(80.0,50.0)[l]{$\bar{D}$}
\ArrowLine(79.0,50.0)(58.0,60.0) 
\Text(47,30)[b] {down quarks, s channel}
\end{picture} \ 
{} \qquad\allowbreak
\begin{picture}(95,79)(0,0)
\Text(15.0,70.0)[r]{$u$}
\ArrowLine(16.0,70.0)(58.0,70.0) 
\Text(80.0,70.0)[l]{$D$}
\ArrowLine(58.0,70.0)(79.0,70.0) 
\Text(54.0,60.0)[r]{$W^+$}
\DashArrowLine(58.0,70.0)(58.0,50.0){3.0} 
\Text(15.0,50.0)[r]{$\bar{u}$}
\ArrowLine(58.0,50.0)(16.0,50.0) 
\Text(80.0,50.0)[l]{$\bar{D}$}
\ArrowLine(79.0,50.0)(58.0,50.0) 
\Text(47,30)[b] {up quarks, t channel}
\end{picture} \ 
}
\vskip-1cm
{
\unitlength=1.0 pt
\SetScale{1.0}
\SetWidth{0.7}      
\scriptsize    
{} \qquad\allowbreak
\begin{picture}(95,79)(0,0)
\Text(15.0,70.0)[r]{$d$}
\ArrowLine(16.0,70.0)(58.0,70.0) 
\Text(80.0,70.0)[l]{$D$}
\ArrowLine(58.0,70.0)(79.0,70.0) 
\Text(57.0,60.0)[r]{$Z$}
\DashLine(58.0,70.0)(58.0,50.0){3.0} 
\Text(15.0,50.0)[r]{$d$}
\ArrowLine(16.0,50.0)(58.0,50.0) 
\Text(80.0,50.0)[l]{$D$}
\ArrowLine(58.0,50.0)(79.0,50.0) 
\Text(47,30)[b] {down quarks, t channel 1}
\end{picture} \ 
}
{
\unitlength=1.0 pt
\SetScale{1.0}
\SetWidth{0.7}      
\scriptsize    
{} \qquad\allowbreak
\begin{picture}(95,79)(0,0)
\Text(15.0,70.0)[r]{$d$}
\ArrowLine(16.0,70.0)(58.0,70.0) 
\Text(80.0,70.0)[l]{$D$}
\ArrowLine(58.0,70.0)(79.0,70.0) 
\Text(57.0,60.0)[r]{$Z$}
\DashLine(58.0,70.0)(58.0,50.0){3.0} 
\Text(15.0,50.0)[r]{$\bar{d}$}
\ArrowLine(58.0,50.0)(16.0,50.0) 
\Text(80.0,50.0)[l]{$\bar{D}$}
\ArrowLine(79.0,50.0)(58.0,50.0) 
\Text(47,30)[b] {down quarks, t channel 2}
\end{picture} \ 
}
{
\unitlength=1.0 pt
\SetScale{1.0}
\SetWidth{0.7}      
\scriptsize    
{} \qquad\allowbreak
\begin{picture}(95,79)(0,0)
\Text(15.0,70.0)[r]{$\bar{d}$}
\ArrowLine(16.0,70.0)(58.0,70.0) 
\Text(80.0,70.0)[l]{$\bar{D}$}
\ArrowLine(58.0,70.0)(79.0,70.0) 
\Text(57.0,60.0)[r]{$Z$}
\DashLine(58.0,70.0)(58.0,50.0){3.0} 
\Text(15.0,50.0)[r]{$\bar{d}$}
\ArrowLine(58.0,50.0)(16.0,50.0) 
\Text(80.0,50.0)[l]{$\bar{D}$}
\ArrowLine(79.0,50.0)(58.0,50.0) 
\Text(47,30)[b] {down quarks, t channel 3}
\end{picture} \ 
}
\end{center}
\end{figure}

\begin{figure}

\caption{The $D\overline{{D}}$ pair total production cross section (solid
line) as a function of $D$ quark mass is shown as calculated with
CompHep (triangles) and with MadGraph (circles). The dashed line is
for the gluon contribution and the dotted line is for the quark contribution.
The two generators agree within 2\% and the effect of PDF uncertainties
is computed to be less than 10\% through 5 orders of magnitude in
the cross section.\label{cap:total-cross-section-signal}}

\begin{center}\includegraphics[%
  scale=0.45]{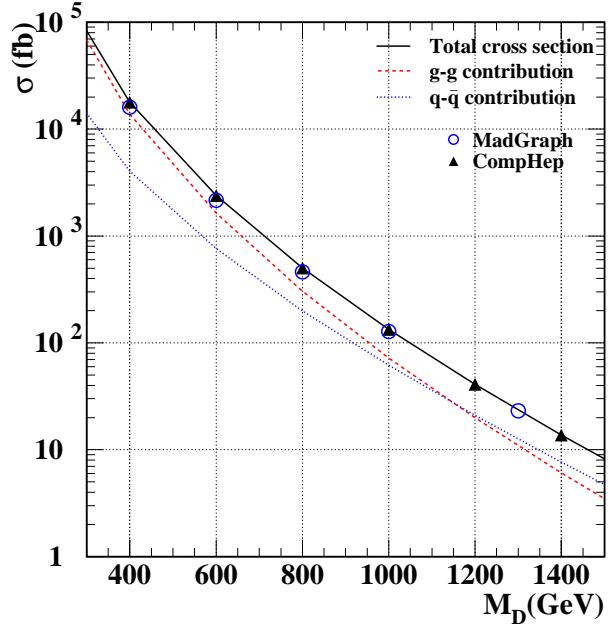}\end{center}
\end{figure}

The main tree level Feynman diagrams for the pair production of $D$
quarks at LHC are presented in Fig.~\ref{cap:signal-feynman}. The
$gD\overline{D}$ and $\gamma D\overline{D}$ vertices are the same
as their SM down quark counterparts. The modification to the $ZD\overline{D}$
vertex due to $d-D$ mixing can be neglected due to the small value
of $\sin\phi$. The Lagrangian in Eq. (\ref{lagrangian}) was implemented
into tree level event generators, \emph{Comphep} \cite{R-calchep}
version 4.3 and \emph{Madgraph} \cite{madgraph} version 2.3~. The
total pair production cross sections from these two Monte Carlo generators
are shown in Fig.~\ref{cap:total-cross-section-signal}. The difference
in cross section calculated by these two generators, the first one,
based on full matrix element calculation and the other on the numerical
methods is less than 5\% for the range of $D$ quark mass from 400
to 1400$\;$GeV. The impact of uncertainties in parton distribution
functions (PDFs) \cite{R-cteq}, is calculated \textcolor{red}{}by
using different PDF sets, to be less than 10\% for the same range.
For example at $m_{D}=800$~GeV and $Q^{2}=m_{Z}^{2}$, the cross
section values are 450 (CompHep, CTEQ6L1) and 468 (CompHep, CTEQ5L)
versus 449 (MadGraph, CTEQ6L1) and 459 (MadGraph, CTEQ5L) fb with
an error of about one percent. For the same PDF set, the two programs
give the same answer validating both of the MC generators against
each other and the implementation of the model. The same figure also
contains the partial ($g$$g$ and $q\overline{q}$) contributions
showing that the largest contribution to the total cross section comes
from the first three diagrams for $D$ quark masses $<$1100$\;$GeV,
while for higher $D$ quark masses, contributions from $s$-channel
$q\bar{q}$ subprocesses becomes dominant. For these computations,
$q\bar{q}$ are assumed to be only from the first quark family since,
the contribution to the total cross section from $s\bar{s}$ is about
10 times smaller and the contribution from $c\bar{c}$ and $b\bar{b}$
are about 100 times smaller. The $t$-channel diagrams mediated by
$Z$ and $W$ bosons, shown on the bottom row of Fig.~\ref{cap:signal-feynman},
which are suppressed by the small value of $\sin\phi$ (for example
0.4 fb at $m_{D}=800$~GeV) are also included in the signal generation.

The isosinglet quarks being too heavy are expected to immediately
decay into SM particles. The decay channels leading to possible discovery
for the $D$ quark pairs are summarized in Table \ref{cap:Possible-signal-decay-channels}.
We have initially focused on the 4 lepton final states of the neutral
channel only: although it has the smallest branching ratio, the possibility
of reconstructing the invariant mass for $Z$ bosons and thus for
both $D$ quarks makes it favorable for a feasibility study. Therefore
the final state we would be looking for is composed of two high transverse
momenta jets and two $Z$ bosons, all coming from the decay of the
$D$ quarks. The high transverse momentum of the jets coming from
the $D$ quark decays can be used to distinguish the signal events
from the background. 

All the SM processes allowed in proton collisions yielding two $Z$
bosons and two jets (originating from any parton except $b$ and $t$
quarks) were considered as background. These can be classified in
three categories via the initial state partons: $qq$, $qg$ and $gg$
where $g$ stands for gluon and $q$ can be any quark or anti-quark
from the first two families. The contribution from third quark family
is assumed to be negligibly small due to mass and PDF arguments. Although
a complete list can be seen in \cite{details}, one should note that
the 78\% of the background cross section originates from the processes
in first two categories : $qg\rightarrow qgZZ$ (where $q=u,\, d,\,\bar{{d}},\, c,\, s,\,\bar{s}$)
contributes 58\%, and $q\bar{{q}}\rightarrow ggZZ$ (where $q=d,\, u,\, s,\, c$)
contributes 20\%. The simple requirements imposed at the generator
level are:\begin{eqnarray}
|\eta_{p}| & < & 2.5\;,\label{eq:generator level cuts}\\
P_{T,p} & > & 100\;\mbox{{GeV}}\;,\nonumber \\
R_{p\overline{p}} & > & 0.4\;,\nonumber \\
|\eta_{Z}| & < & 5.0\nonumber \end{eqnarray}
 where R is the cone separation angle between two partons ($p=$$d$,$\overline{d})$,
$\eta_{p}$ and $\eta_{Z}$ are pseudorapidities of a parton and $Z$
boson respectively, and $P_{T,p}$ is the parton transverse momentum.
The selection of the $\eta$ region is driven by partonic spectra
pseudorapitidy distributions, which are peaked in the barrel detector
area. The signal cross section was calculated with both generators
as a function of the $D$ quark mass, but the background only with
Madgraph as it is faster in numerical evaluation. For $m_{D}=800$~GeV,
using the generator level cuts listed in Eq. (\ref{eq:generator level cuts}),
the cross section of the $ZZ\,2jet$ channel is found to be $\sigma_{signal}=45.4\pm2$~fb~
(CompHep) whereas the SM background for the same final state particles
is $\sigma_{bg}=345\pm17$~fb~ (MadGraph). Already at this level,
a significance of $S/\sqrt{B}\approx2.4$ can be obtained with an
integrated luminosity of 1 fb$^{-1}$.

\begin{table}

\caption{The promising signal channels. The fourth column contains the branching
ratios of the SM particles, whereas the last column has the total
branching ratios.\label{cap:Possible-signal-decay-channels}}

\begin{center}{\small }\begin{tabular}{|c|c|c|c|c|}
\hline 
{\footnotesize $D\bar{{D}}\rightarrow$}&
{\footnotesize Final State}&
{\footnotesize Expected Signal }&
{\footnotesize Decay B.R.}&
{\footnotesize Total B.R.}\tabularnewline
\hline
\hline 
{\footnotesize $Z\, Z\, d\,\bar{{d}}$}&
{\footnotesize $Z\rightarrow l\bar{{l}}$ $Z\rightarrow l\bar{{l}}$ }&
{\footnotesize $4\: l\:+2\: jet$}&
{\footnotesize $0.07\times0.07$}&
{\footnotesize $0.0005$}\tabularnewline
\cline{2-2} \cline{3-3} \cline{4-4} \cline{5-5} 
\multicolumn{1}{|c|}{{\footnotesize $0.33\times0.33$}}&
{\footnotesize $Z\rightarrow l\bar{{l}}$ $Z\rightarrow v{v}$ }&
{\footnotesize $2\: l\:+2\: jet+E\!\!\!/_{T}$}&
{\footnotesize $2\times0.07\times0.2$}&
{\footnotesize $0.0028$}\tabularnewline
\cline{2-2} \cline{3-3} \cline{4-4} \cline{5-5} 
\multicolumn{1}{|c|}{}&
{\footnotesize $Z\rightarrow l\bar{{l}}$ $Z\rightarrow q\bar{{q}}$ }&
\multicolumn{1}{c|}{{\footnotesize $2\: l\:+4\: jet$}}&
{\footnotesize $2\times0.07\times0.7$}&
{\footnotesize $0.0107$}\tabularnewline
\hline 
{\footnotesize $Z\, W\, d\, u$ }&
{\footnotesize $Z\rightarrow l\bar{{l}}$ $W\rightarrow l\bar{{v}}$ }&
{\footnotesize $3\: l\:+2\: jet+E\!\!\!/_{T}$}&
{\footnotesize $0.07\times0.21$}&
{\footnotesize $0.0065$}\tabularnewline
\cline{2-2} \cline{3-3} \cline{4-4} \cline{5-5} 
\multicolumn{1}{|c|}{{\footnotesize $2\times0.33\times0.67$}}&
{\footnotesize $Z\rightarrow l\bar{{l}}$ $W\rightarrow q\bar{{q}}$ }&
{\footnotesize $2\: l\:+4\: jet$}&
{\footnotesize $0.07\times0.68$}&
{\footnotesize $0.0211$}\tabularnewline
\hline
\end{tabular}\end{center}
\end{table}

\section{Observation in ATLAS detector using $4l\;2jet$ channel}

Using the preselection cuts listed in the previous section, 5000 signal
events at $m_{D}=800$~GeV in CompHep and slightly relaxing the jet
transverse momentum cut, ($P_{T,j}>50$~GeV), 40000 background events
in MadGraph were generated. The $D$ quarks in signal events were
made to decay in CompHep into SM particles. The final state particles
for both signal and background events were fed into PYTHIA version
6.218 \cite{R-pythia} for initial and final state radiation, as well
as, hadronization using the CompHep-Pythia and MadGraph-Pythia interfaces
provided by Athena (the ATLAS offline software framework) v9.0.3.
To incorporate the detector effects, all event samples were processed
through the ATLAS fast simulation tool, ATLFAST \cite{R-AtlFast},
and the final analysis has been done using physics objects that it
produced. 

It must be noted that ATLFAST uses a parameterization for electrons,
muons and hadrons without the detailed simulation of showers in the
calorimeters. There is also a separate parameterization on the resolution
for muons and electron tracks for the inner detector efficiencies.
Minimum transverse energy of electromagnetic and hadronic clusters
to be considered as electron or jet showers are $E_{T}>5$~GeV and
$E_{T}>10$~GeV, respectively. Electromagnetic and hadronic cells
in ATLFAST have the same granularity : $\Delta\eta\times\Delta\phi=0.1\times0.1$.
The electron isolation criteria requires a minimum distance $\Delta R\equiv\sqrt{(\Delta\eta)^{2}\times(\Delta\phi)^{2}}\ge0.4$
from other clusters and a maximum transverse energy deposition, $E_{T}<10$~GeV
in outer cells accompanying the electron candidate. These outer cells
are to be in a cone of radius $\Delta R=0.2$ along the direction
of emission. The jets are reconstructed using the cone algorithm with
the $\Delta R=0.4$ cone size. The smearing of particle clusters and
jets is tuned to what is expected for the performance of the ATLAS
detector from full simulation and reconstruction using the GEANT package\cite{Geant}.

For this initial feasibility study, where the aim is to reconstruct
the invariant mass of both $D$ quarks, only $e$ and $\mu$ decays
of $Z$ bosons are considered. Although the effective cross section
becomes small compared to other decay channels, the benefits of this
selection for a clean signal and for correct invariant mass reconstruction
are indisputable. Other studies involving the invisible decays of
$Z$ boson and leptonic decays of $W$ boson are in preparation. For
the initial state particles, $gg$, $u\bar{{u}}$ and $d\bar{{d}}$
sub-channels were studied separately, and for final state particles
only light quark jets were considered, all using CTEQ6L1 as PDF \cite{R-cteq}.
For the case of $M_{D}=800$~GeV, the contributions from each sub-channel
to the final cross section were about 50 \%, 32\% and 18\% respectively.

\subsection{$4l\;2jet$ channel, both $Z\rightarrow\mu\mu$ case}

\begin{figure}

\caption{The transverse momentum cuts for muons (upper set) and jets (lower
set). The plots for signal at $M_{D}=800$~GeV and the SM backgrounds
are shown with arrows pointing at the cut values. The highest jet
$P_{T}$ for signal events peaks around 300~GeV, whereas for background
no such peak is observed.\label{fig:Transverse-momentum-cuts-mumu}}

\begin{center}\includegraphics[%
  scale=0.61]{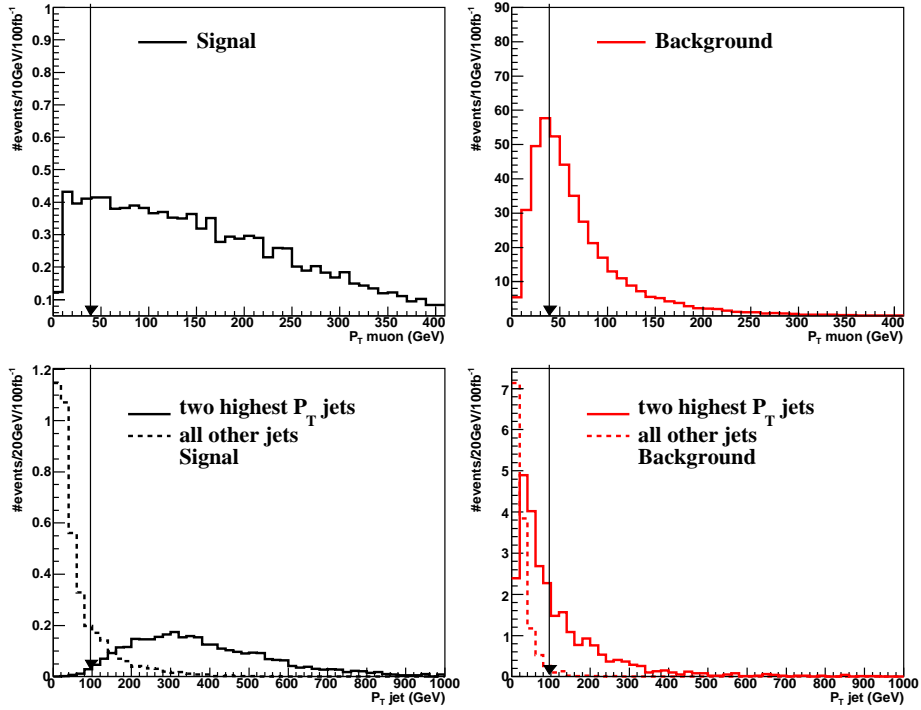}\end{center}
\end{figure}
\begin{figure}

\caption{The invariant mass plots for signal at $M_{D}=800$~GeV and background
events in the $Z\rightarrow\mu\mu$ case, obtained from two reconstructed
$Z$ bosons and two highest $P_{T}$ jets. \label{fig:invarient-mass-plots-mumu}}

\begin{center}\includegraphics[%
  scale=0.6]{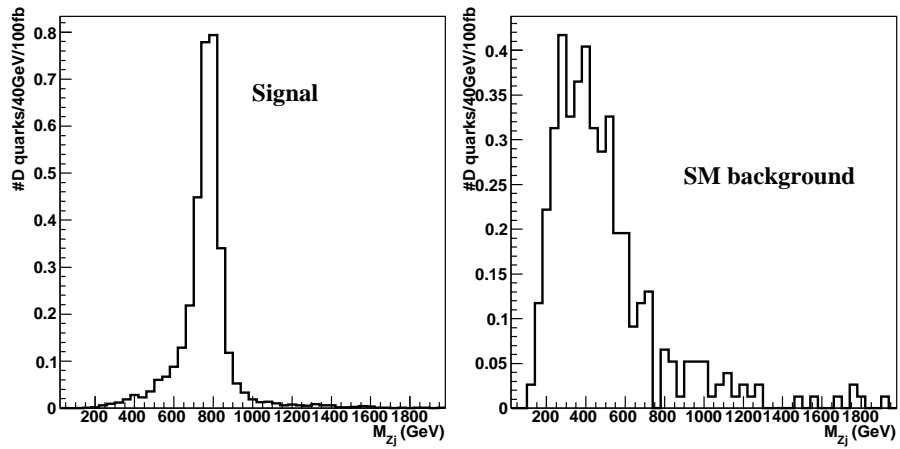}\end{center}
\end{figure}

To reconstruct the two $Z$ bosons, four isolated muons were required.
The efficiency of finding 4 isolated muons is roughly 50\%. Since
it is not known which two muons were originating from the same $Z$
boson, all three possible combinations (assuming the lepton charge
is not measured) were considered. The reconstructed invariant mass
of the muon pairs was required to be close to the $Z$ boson mass:
within a window of 20~GeV around the central value of 90~GeV. If
all pair combinations happened to be within the mass window, the one
with the invariant mass closest to the measured value of $Z$ boson
mass was taken. The efficiency of reconstructing two $Z$ bosons was
about 90\%. High transverse momentum cuts as in equation set below
were applied to all leptons and jets to distinguish the signal events
from background. One should note that, all the muons in such an event
would satisfy the currently set trigger conditions for leptons $P_{T,lepton}\geq20$~GeV
\cite{tdaq-tdr}. The momentum distributions for all muons in an event
and the imposed cuts for both signal and background can be seen in
Fig.~\ref{fig:Transverse-momentum-cuts-mumu} upper set where the
vertical arrow points at the imposed cut value. The percentage of
muons surviving the transverse momentum cut in equation (\ref{eq:mu analysis cuts})
is about 60\%. The events with at least two jets with transverse momentum
greater than 100~GeV were kept in order to fully reconstruct the
invariant mass of $D$ quarks. In the lower set of Fig.~\ref{fig:Transverse-momentum-cuts-mumu},
the solid line shows the momenta of the two most energetic jets in
an event, whereas the dashed curve is for all other jets in the same
event, again the vertical arrows pointing at the cut values. The two
most energetic jets and the two previously reconstructed $Z$ bosons
were combined to reconstruct the $D$ quark pair which was the goal
of this study. However, the correct association of the two most energetic
jets to the two reconstructed $Z$ bosons is not known and involves
combinatorics. Since the mass of the $D$ quark is not known a priori,
the {}``mass window around the central value'' method that was used
for $Z$ boson reconstruction cannot be applied. The effect of wrong
jet-$Z$ association has the impact of enlarging the tails for the
signal invariant mass distribution. This problem was partially solved
by selecting the combination with the smallest absolute value of the
difference between the two reconstructed $D$ quark masses. Reconstructed
invariant mass histograms are given separately in Fig.~\ref{fig:invarient-mass-plots-mumu}
for both signal and background cases. Therefore, the list of all the
analysis level cuts becomes:

\begin{eqnarray}
N_{\mu} & = & 4\;,\nonumber \\
P_{T,\mu} & > & 40\;\mbox{{GeV}}\;,\label{eq:mu analysis cuts}\\
M_{Z} & = & 90\;±\;20\;\mbox{{GeV}}\;,\nonumber \\
N_{jet} & \geq & 2\:,\nonumber \\
P_{T,jet} & \geq & 100\;\mbox{{GeV}}\;.\nonumber \end{eqnarray}

The selection cut efficiencies for both signal and background events
are given in Table \ref{cap:subchannel-efficiencies-mumu} where each
efficiency is calculated relative to the previous one, and the last
column contains the combined efficiency value. 

\begin{table}[!h]

\caption{The individual selection cut efficiencies in percent for the both
$Z\rightarrow\mu\mu$ case, for signal and background}

\begin{center}\begin{tabular}{|c|c|c|c|c|c|c|}
\hline 
channel&
$N_{\mu}$&
$M_{Z}$&
$P_{T,\mu}$&
$N_{jet}$&
$P_{T,jet}$&
$\epsilon_{combined}$\tabularnewline
\hline
\hline 
Signal&
48&
91&
59&
100&
95&
25\tabularnewline
\hline 
Background&
34&
96&
16&
96&
12&
0.6\tabularnewline
\hline
\end{tabular}\label{cap:subchannel-efficiencies-mumu}\end{center}
\end{table}

\subsection{$4l\;2jet$ channel, both $Z\rightarrow ee$ case}

\begin{figure}

\caption{The $D$ quark invariant mass reconstruction in $Z\rightarrow ee$
case for all signal sub-channels as compared to SM background. The
signal was generated for $M_{D}=800$~GeV. \label{fig:ee-invarient-mass} }

\begin{center}\includegraphics[%
  scale=0.6]{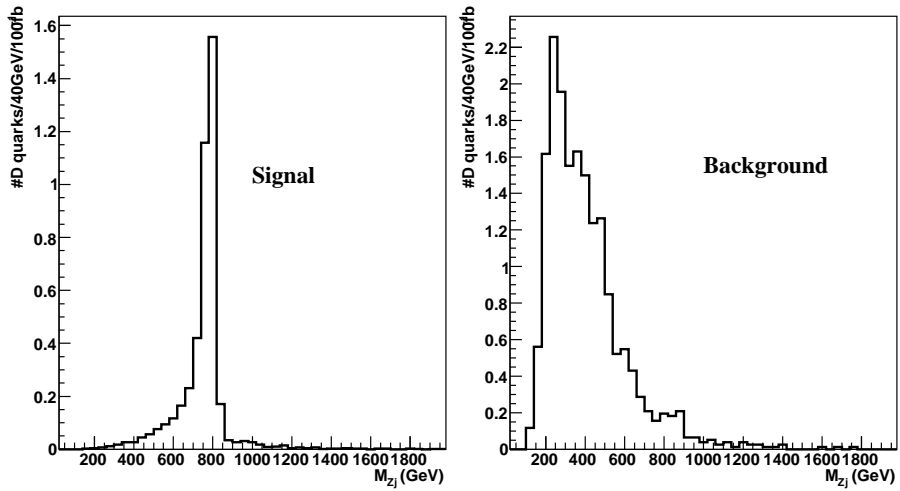}\end{center}
\end{figure}

The cuts for this channel are same as for the channel above (Eq.~(\ref{eq:mu analysis cuts})),
except the cut on the transverse momentum of electron, $P_{T,e}$$>$15
GeV has been used. The existence of at least one electron originating
from the $Z$ boson decay which satisfies the isolated lepton trigger
condition has also been checked for all 4 electron events. The method
for reconstructing the $Z$ boson and eventually the $D$ quark remains
the same as in the previous case. The selection cut efficiencies for
this case are shown in Table~\ref{tab:ee-cut-efficiencies} for both
signal and background. The electron cuts leave slightly higher surviving
events for both signal and background as compared to the muons. This
difference and the different cut values for lepton transverse momenta
can be attributed to the over optimistic electron reconstruction in
the ATLAS software. The impact of this issue, which was under investigation
at the time of this writing, was estimated by comparing the fits to
the reconstructed $Z$ bosons from both electrons and muons and was
found to be about 5\%. The invariant mass spectra is given in Fig.~\ref{fig:ee-invarient-mass}
for a bin width of 40 GeV, showing the expected number of signal events
being larger than the background ones. 

\begin{table}[!h]

\caption{The individual selection cut efficiencies in percent for the both
$Z\rightarrow ee$ case for both signal and background. \label{tab:ee-cut-efficiencies}}

\begin{center}\begin{tabular}{|c|c|c|c|c|c||c|}
\hline 
channel&
$N_{e}$&
$M_{Z}$&
$P_{T,e}$&
$N_{jet}$&
$P_{T,jet}$&
$\epsilon_{combined}$\tabularnewline
\hline
\hline 
Signal&
40&
99&
87&
100&
95&
33\tabularnewline
\hline 
Background&
37&
98&
83&
94&
7&
2.0\tabularnewline
\hline
\end{tabular}\end{center}
\end{table}

\subsection{$4l\;2jet$ channel, one $Z\rightarrow ee$ and one $Z\rightarrow\mu\mu$
case}

\begin{figure}

\caption{The $D$ quark invariant mass reconstruction in $ZZ\rightarrow ee\mu\mu$
case for all signal sub-channels as compared to SM background. The
signal was generated for $M_{D}=800$~GeV. \label{fig:emu-invarient-mass}}

\begin{center}\includegraphics[%
  scale=0.6]{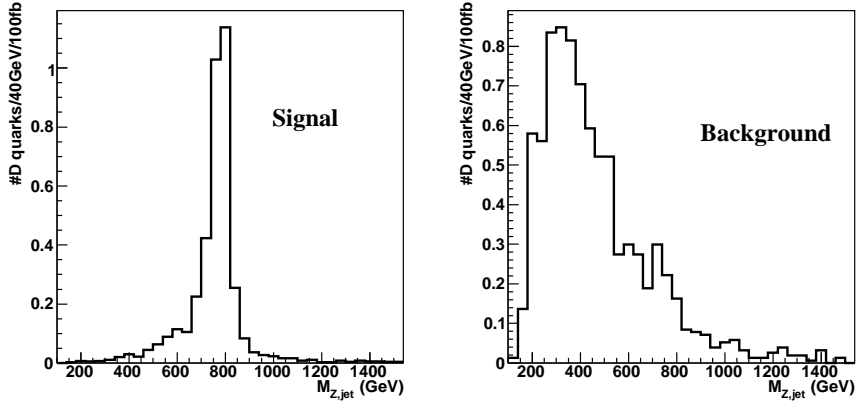}\end{center}
\end{figure}

This case is based on two isolated electrons, two isolated muons and
two jets. There is no ambiguity in the lepton selection for $Z$ invariant
mass reconstruction thus a simpler reconstruction algorithm suffices.
Both muons and at least one electron satisfy the trigger condition.
The event selection cuts are summarized in the equation set \ref{eq:emu-analysis-cuts}:

\begin{eqnarray}
N_{\mu} & = & 2\;,\quad N_{e}=2\;,\label{eq:emu-analysis-cuts}\\
P_{T,\mu} & > & 40\;\mbox{{GeV}}\;,\nonumber \\
P_{T,e} & > & 15\;\mbox{{GeV}}\;,\nonumber \\
N_{jet} & \geq & 2\nonumber \\
P_{T,jet} & \geq & 100\;\mbox{{GeV}}\;,\nonumber \\
M_{Z} & = & 90\;±\;20\;\mbox{{GeV}}\;,\nonumber \end{eqnarray}

The selection cut efficiencies are given in Table~\ref{tab:emu-cut-efficiencies}.
Since the branching ratio is higher by a factor of two, compared to
the first and second cases, this case yields more signal events and
dominates the results. The reconstructed invariant mass for the signal
and the SM background is given in Fig.~\ref{fig:emu-invarient-mass}
showing that the expected number of signal events is higher than for
the background, in the region of the peak.

\begin{table}[!h]

\caption{The individual selection cut efficiencies for one $Z\rightarrow ee$
and one $Z\rightarrow\mu\mu$ sub-case. The subscript $l$ represents
both electron and muon cases. \label{tab:emu-cut-efficiencies}}

\begin{center}\begin{tabular}{|c|c|c|c|c|c|c|}
\hline 
channel&
$N_{l}$&
$M_{Z}$&
$P_{T,l}$&
$N_{jet}$&
$P_{T,jet}$&
$\epsilon_{combined}$\tabularnewline
\hline
\hline 
Signal&
44&
94&
71&
100&
93&
28\tabularnewline
\hline 
Background&
35&
97&
34&
95&
10&
1.1\tabularnewline
\hline
\end{tabular}\end{center}
\end{table}

\section{Results }

\begin{figure}[h]

\caption{Combined results for possible signal observation at $M_{D}=$ 600,
800, 1000, 1200~GeV . The reconstructed $D$ quark mass and the relevant
SM background are plotted for a luminosity of 100 fb$^{-1}$ which
corresponds to one year of nominal LHC operation. The dark line shows
the signal and background added, the dashed line is for signal only
and the light line shows the SM background. \label{fig:Combined-results}}

\begin{spacing}{0}
\noindent \begin{center}\includegraphics[%
  clip,
  scale=0.47]{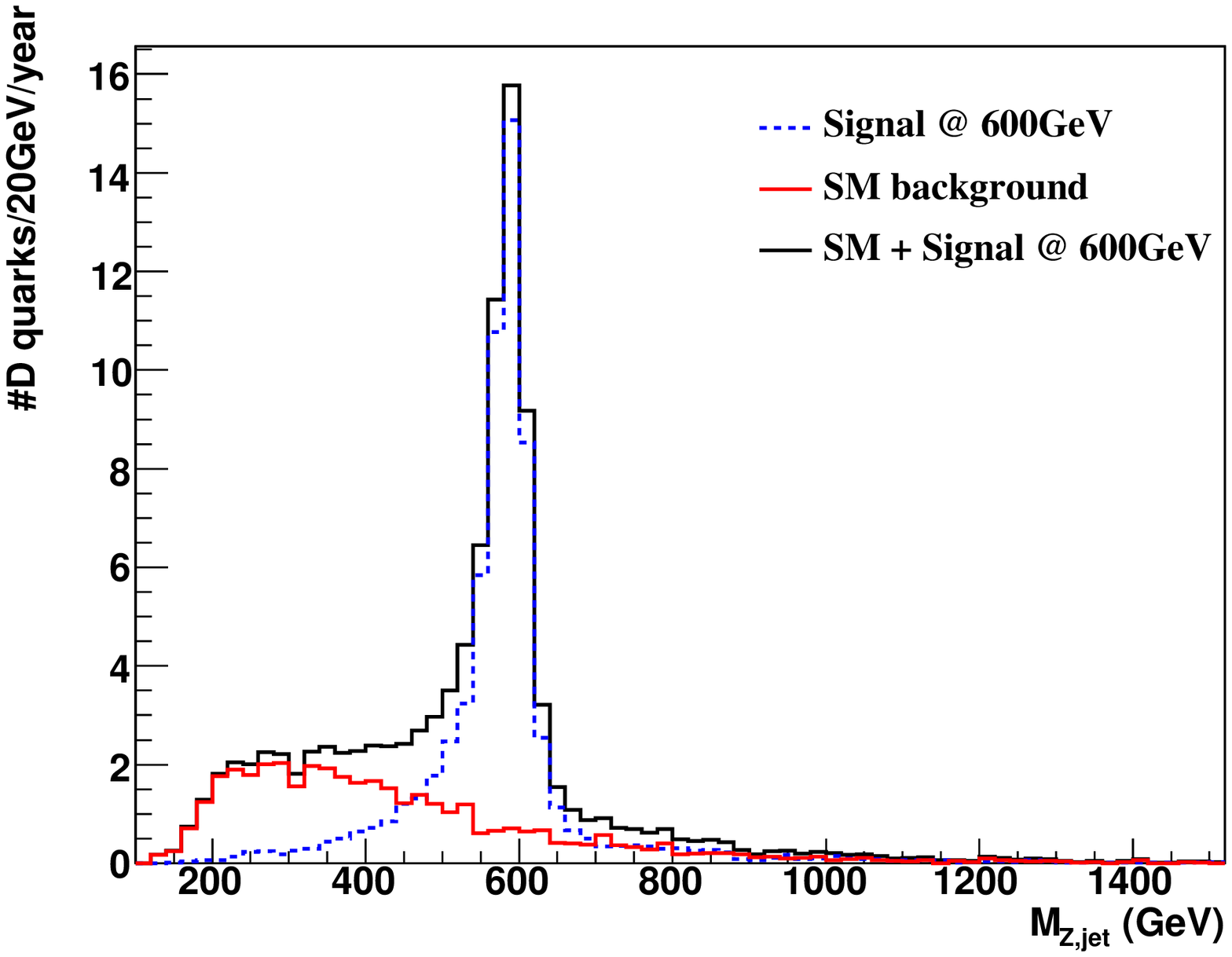}\includegraphics[%
  scale=0.47]{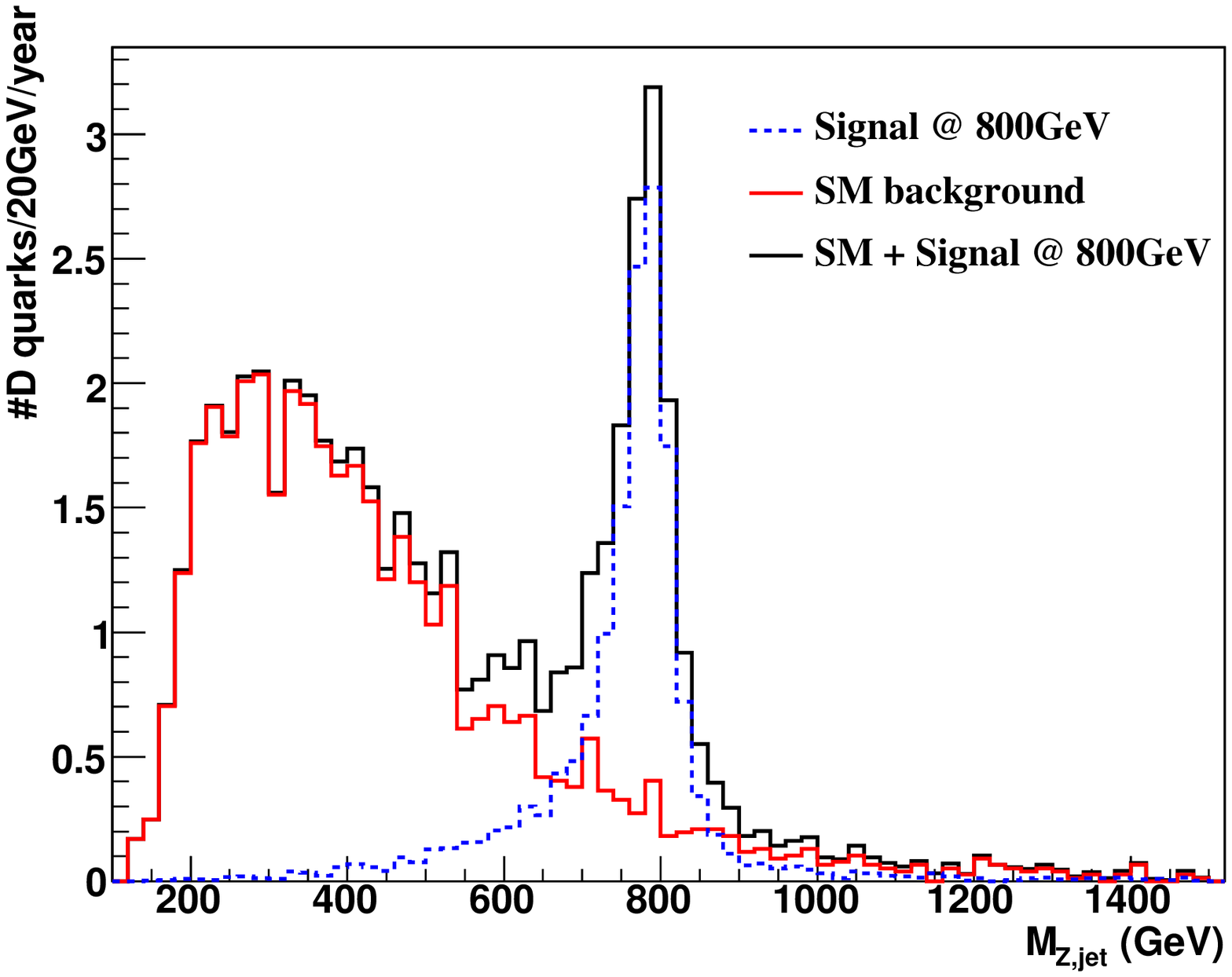}\vskip-1cm\end{center}
\end{spacing}

\noindent \begin{center}\includegraphics[%
  scale=0.47]{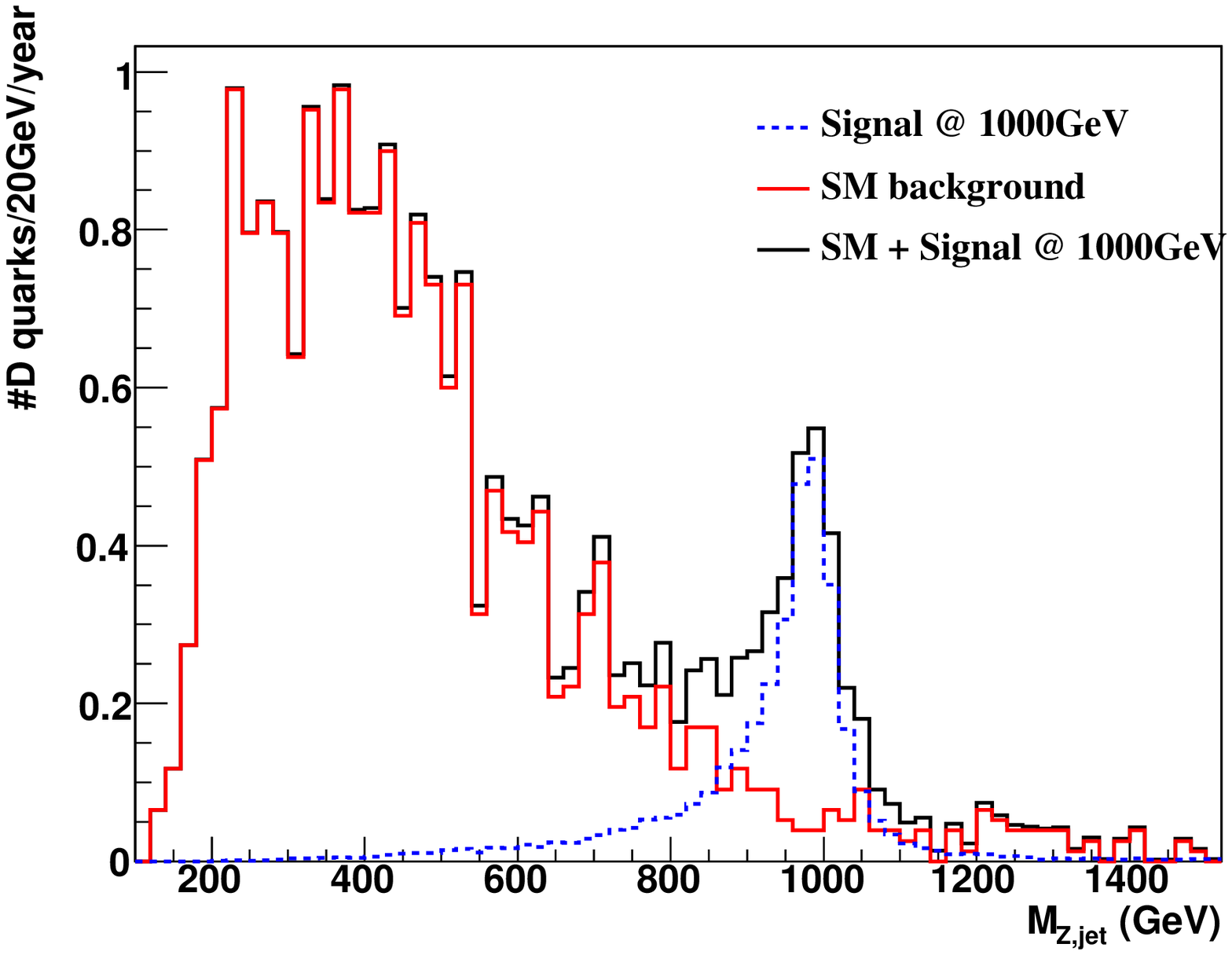}\includegraphics[%
  scale=0.47]{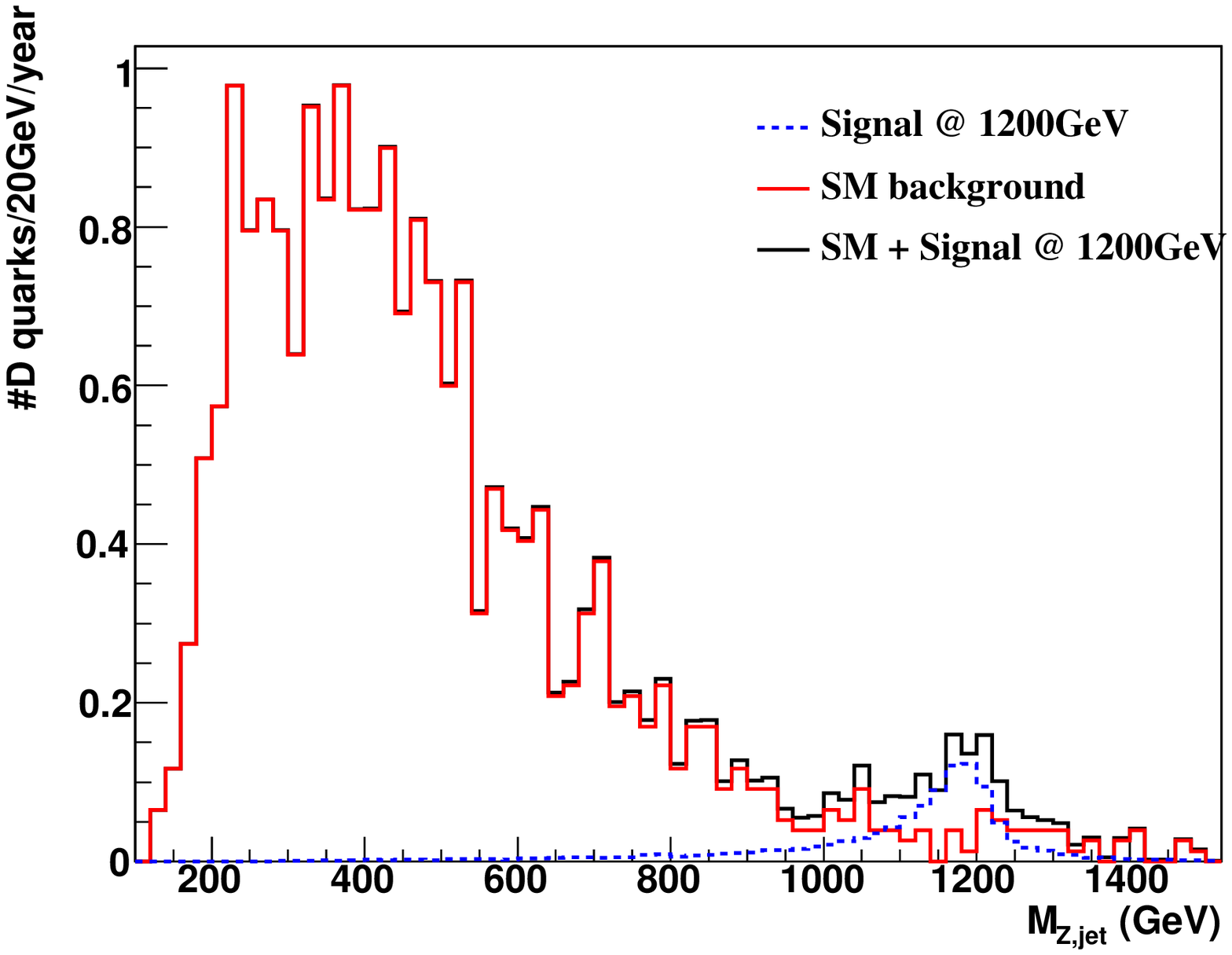}\end{center}
\end{figure}

All three above mentioned leptonic reconstruction cases were also
considered for systematic studies at other $D$ quark mass values:$M_{D}=$
600, 1000, 1200~GeV (The details of this analysis can be found in
\cite{details}). For values of $D$ quark mass larger than 900~GeV,
in order to increase the expected statistical significance of signal
identification, the cuts on the basic kinematic variables were modified
as: \begin{eqnarray}
P_{T,\mu} & > & 50\;\mbox{{GeV}}\;,\\
P_{T,e} & > & 20\;\mbox{{GeV}}\;,\nonumber \\
P_{T,jet} & > & 120\;\mbox{{GeV}}\;.\nonumber \end{eqnarray}
Using the convention of defining a running accelerator year as $1\times10^{7}$seconds,
one LHC year at the full design luminosity corresponds to 100~$fb^{-1}$.
For one such year worth of data, all the signal events are summed
and compared to all SM background events as shown in Fig.~\ref{fig:Combined-results}.
It is evident that for the lowest of the considered masses, the studied
channel gives an easy detection possibility, whereas for the highest
mass case ($M_{D}$=1200 GeV) the signal to background ratio is of
the order of unity. If Nature has assigned a high mass to the $D$
quark, other possible detection channels would be either to tag one
$Z$ via its leptonic decays and consider the neutrino decays of the
other, or would involve hadronic decays of at least one $Z$ and methods
to disentangle the jet association. The detailed study of these modes
as well as the charged current decay modes is deferred to future work. 

\begin{figure}

\caption{On the left: the expected statistical significance after 3 years
of running at nominal LHC luminosity assuming Gaussian statistics.
The vertical line shows the limit at which the event yield drops below
10 events. On the right: the integrated luminosities for 3 sigma observation
and 5 sigma discovery cases as a function of $D$ quark mass. The
bands represent statistical uncertainties originating from finite
MC sample size. \label{cap:Statistical-significance-NC}}

\begin{center}\includegraphics[%
  bb=0bp 0bp 520bp 520bp,
  scale=0.49]{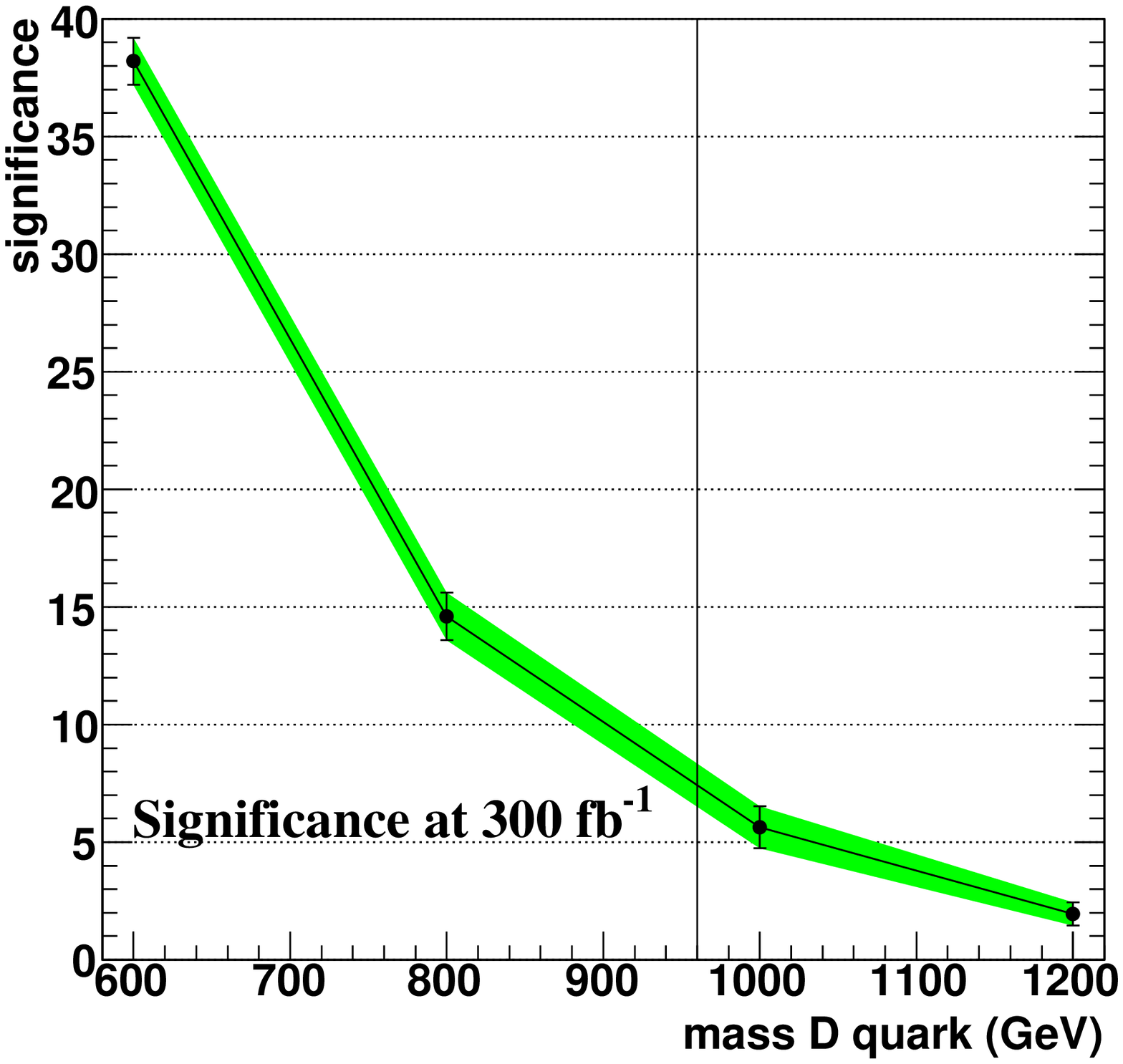}\includegraphics[%
  scale=0.46]{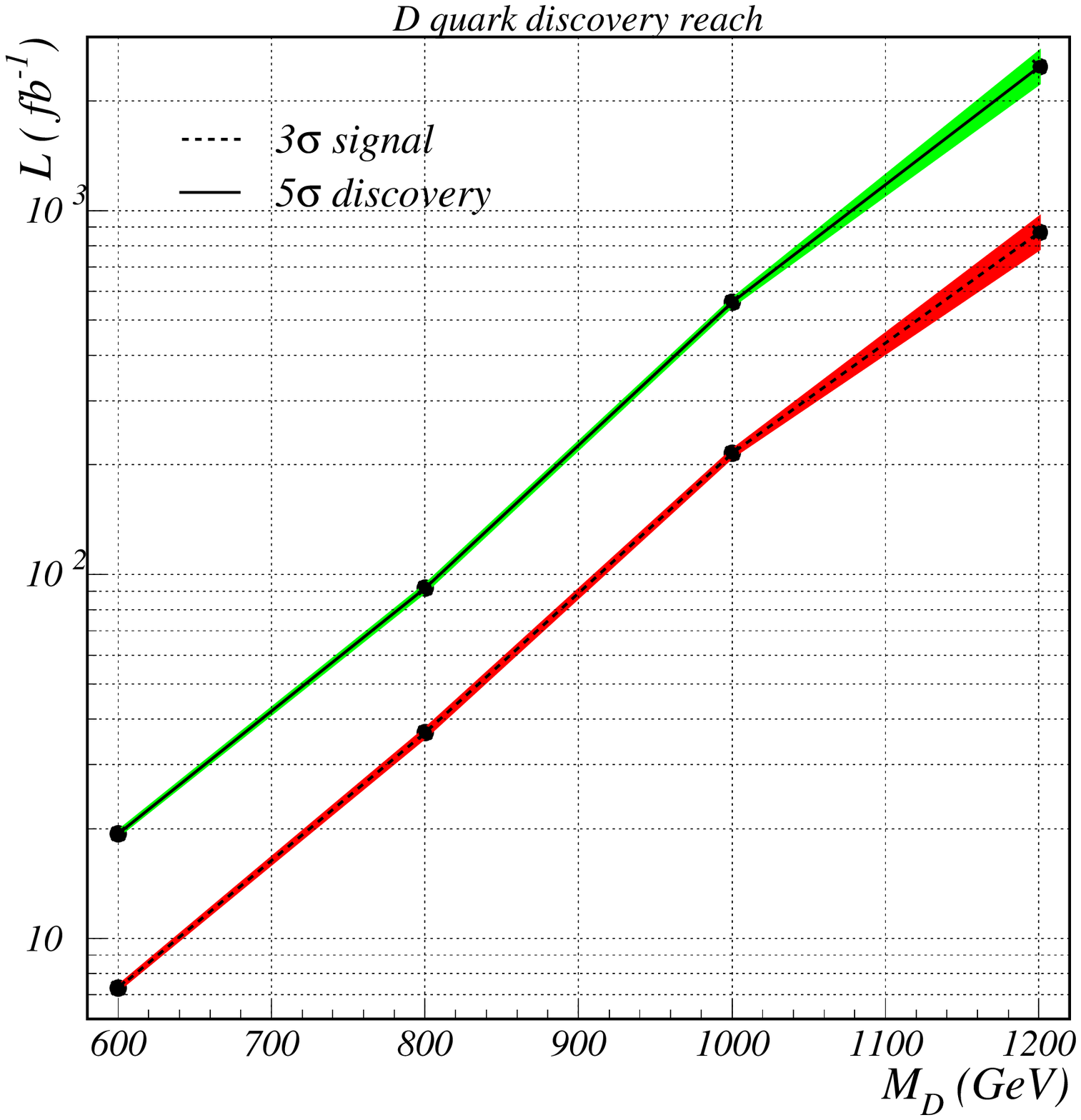}\end{center}
\end{figure}

For each $D$ quark mass value that was considered, a Gaussian is
fitted to the invariant mass distribution around the $D$ signal peak
and a polynomial to the background invariant mass distribution. The
number of accepted signal ($S$) and background ($B$) events are
integrated using the fitted functions in a mass window whose width
is equal to $2\sigma$ around the central value of the fitted Gaussian.
The significance is then calculated at each mass value as $S/\sqrt{B}$,
using the number of integrated events in the respective mass windows.
The expected signal significance for three years of nominal LHC luminosity
running is shown in Fig. \ref{cap:Statistical-significance-NC} left
hand side. The shaded band in the same plot represents the statistical
errors originating from the fact that for each signal mass value,
a finite number of Monte Carlo events was generated at the start of
the analysis and the surviving events were selected from this event
pool. The statistical errors were calculated using binomial distribution
and propagated to all significance and luminosity calculations. Given
the small value of event yield, the Poisson probability distribution
is more appropriate to set the observation and discovery limits \cite{R-statistics book}.
Following the PDG, the Gaussian probabilities for observing 2, 3 and
5 standard deviations when no signal is expected are defined as 4.55\%,
0.27\% and 5.7$\times10^{-5}$\% respectively. Since the number of
observed events in Poisson distribution (and also in experiment) is
an integer, the values approaching their Gaussian counterparts from
the lower side are taken. Consequently, this procedure always yields
more pessimistic luminosity values. For example at $M_{D}=800$~GeV,
the probability utilized for 3$\sigma$ limit is $1.34\times10^{-3}$,
about half of its Gaussian counterpart. Therefore, the reach of ATLAS
to either exclude the existence of or discover the $D$ quark using
the studied dilepton channel is given in Table~\ref{table:discovery-Lumi}
for different $D$ quark mass values. We observe that for $M_{D}=600$~GeV,
ATLAS could observe the $D$ quark with a significance more than 3
sigma before the end of the first year low luminosity running (10~$fb^{-1}$/year)
whereas to claim discovery with 5 sigma significance, it would need
about 2 years of running time at low luminosity. For $M_{D}=1000$~GeV,
about two years of high luminosity running is necessary for a 3 sigma
signal observation claim. The graphical representation is also given
in Fig.\ref{cap:Statistical-significance-NC} right hand side. 

For these results, a possible source of systematic error is the selection
of the QCD scale: although in a complete computation using all possible
diagrams including loops this selection becomes irrelevant, this work
relies on tree level calculations. An increase in QCD scale from the
presently used value of $Z$ boson mass up to $D$ quark mass would
mean a decrease in the total cross sections of about 65\%. \textbf{}Such
a change would also affect the results presented in this work, taking
the 3 sigma signal observation limit mass from 1000~GeV down to 840~GeV
for two years of high luminosity running. It should also be noted
that the detector related backgrounds and the pile-up effects which
could be important at high luminosity LHC, are not considered in this
work.

\begin{table}[!h]

\caption{The required integrated luminosity in fb$^{-1}$ to discover $D$
quark as a function of its mass is shown. Expected signal and background
event number are also given for one year of high luminosity running.\label{table:discovery-Lumi}}

\begin{center}\begin{tabular}{|c|c|c|c|c|}
\hline 
$D$ quark mass (GeV)&
600&
800&
1000&
1200\tabularnewline
\hline
\hline 
signal \& background events / year&
\begin{tabular}{c|c}
38.2&
3.0\tabularnewline
\end{tabular}&
\begin{tabular}{c|c}
9.6&
1.3\tabularnewline
\end{tabular}&
\begin{tabular}{c|c}
2.0&
0.4\tabularnewline
\end{tabular}&
\begin{tabular}{c|c}
0.6&
0.2\tabularnewline
\end{tabular}\tabularnewline
\hline 
Luminosity for 2 $\sigma$ signal (fb$^{-1}$) &
2.4&
18.4&
86&
373\tabularnewline
\hline 
Luminosity for 3 $\sigma$ signal (fb$^{-1}$) &
7.3&
36.7&
215&
870\tabularnewline
\hline 
Luminosity for 5 $\sigma$ signal (fb$^{-1}$) &
19.4&
91.7&
559&
2480\tabularnewline
\hline
\end{tabular}\end{center}
\end{table}

\section{Conclusions}

This study shows that, for a range of $D$ quark mass from 600 to
1000~GeV, ATLAS has a strong potential to find new physics related
to $E_{6}$ GUT. After three years of design luminosity running, the
5 sigma discovery reach for $D$ quark mass is about 920~GeV . The
impact of the assumptions about $D$ quark mass hierarchy and inter-family
mixing is minimal: If the $S$ quark is the lightest or the $D$ quark
mixes mostly to the second family, all the results and conclusions
stay as they are. If the third family is involved, either $B$ quark
being the lightest or through large mixings to the third quark family,
the results would also remain valid, provided no distinction of third
family quarks is imposed. However, if the $b$ quarks are to be identified
to disentangle the $D$ quark mixing, the $b-jet$ tagging efficiencies
should be convoluted with the presented results which would reduce
the number of expected signal and background events by at least 50\%.
To enlarge the experimental reach window for the higher $D$ quark
mass values, the inclusion of other channels (remaining lines of the
Table \ref{cap:Possible-signal-decay-channels}) is also envisaged.
Furthermore, the inclusion of additional gauge bosons predicted by
the $E_{6}$ group could enhance the signal in the $s$ channel if
they have suitable masses. These studies are in preparation.

\subsection*{Acknowledgments}

The authors would like to thank Louis Tremblet and CERN Micro Club
for kindly providing computational facilities. We are grateful to
G. Azuelos, J. D. Bjorken, T. Carli and J. L. Rosner for useful discussions.
R.M. would like to thank Alexander Belyaev for his assistance in model
calculations. R.M. also thanks NSERC/Canada for their support. S.S
and M.Y. acknowledge the support from the Turkish State Planning Committee
under the contract DPT2002K-120250. G.U.'s work is supported in part
by U.S. Department of Energy Grant DE FG0291ER40679. This work has
been performed within the ATLAS Collaboration with the help of the
simulation framework and tools which are the results of the collaboration-wide
efforts.


\begin{thebibliography}{20}
\bibitem{R-atlas-tdr}ATLAS Detector and Physics Performance Technical Design Report. CERN/LHCC/99-14/15.
\bibitem{R-CMS-tdr}CMS collaboration, Technical proposal, CERN-LHCC-94-38.
\bibitem{R-Classification}P. H. Frampton, P. Q. Hung and M. Sher, Phys Rep. \textbf{330}, \emph{263}
(2000).
\bibitem{R-democracy}S. Sultansoy, hep-ph/0004271
\bibitem{LittleHiggs}N. Arkani-Hamed, A. G. Cohen and H. Georgi. Phys. Lett. B \textbf{513},
\emph{232} (2001)., Azuelos, G. et al., Eur. Phys. J. C \textbf{39S2},
\emph{13} (2005).
\bibitem{R-e6}F. Gursey, P. Ramond and P. Sikivie, Phys. Lett. B \textbf{60,} \emph{177}
(1976); F. Gursey and M. Serdaroglu, Lett. Nuovo Cimento \textbf{21},
\emph{28} (1978).
\bibitem{R-String}J.H. Schwarz, Lett. Math. Phys. \textbf{34} \emph{309} (1995); E.
Witten, Nucl. Phys. B \textbf{443} \emph{85} (1995) . 
\bibitem{R-hewet-rizzo}J. Hewett and T. Rizzo, Phys. Rep. \textbf{183}, \emph{193} (1989). 
\bibitem{R-4thfam}E. Arik et al., Phys.Rev.D \textbf{58}, \emph{117701} (1998).
\bibitem{R-aguilar}J.A. Aguilar-Saavedra, Phys.Lett.B \textbf{625}, \emph{234} (2005). 
\bibitem{PDG}S. Eidelmann et. al., P. Phys. Lett. B \textbf{592,} \emph{1} (2004).
\bibitem{Rosner}T. C. Andre and C.L. Rosner, Phys. Rev. D. \textbf{69}, \emph{035009},
(2004). 
\bibitem{N-params}Z. R. Babaev, V.S. Zamiralov and S.F. Sultanov, IHEP preprint, \emph{81-88},
Serpukhov, (1981).
\bibitem{R-e6-orhan-metin}O. Cakir and M. Yilmaz, Europhys. Lett. \textbf{38,} \emph{13} (1997).
\bibitem{e6-higgs}G. Unel and S. Sultansoy, in preparation.
\bibitem{R-calchep}A. Pukhov, {[}arXiv:hep-ph/0412191{]}; E. Boos et al. {[}CompHEP Collaboration{]},
Nucl. Instrum. Meth. A \textbf{534}, \emph{250} (2004).
\bibitem{madgraph}T. Stelzer and W. F. Long, Phys. Commun. \textbf{81}, 357 (1994).
\bibitem{R-cteq}J. Pumplin, D.R. Stump, J. Huston, H.L. Lai, P. Nadolsky and W.K.
Tung, JHEP \textbf{0207}, 012 (2002) {[}arXiv:hep-ph/0201195{]}.
\bibitem{tdaq-tdr}ATLAS Collaboration, Trigger and Data Acquisition Technical Design
Report, LHCC-2003-022/TDR-016.
\bibitem{details}R. Mehdiyev et al., ATL-PHYS-PUB-2005-021 (2005).
\bibitem{R-pythia}T. Sjostrand et al., Computer Phys. Commun. \textbf{135} (2001) 238
(LU TP 00-30, {[}hep-ph/0010017{]})
\bibitem{R-AtlFast}E. Richter-Was et al., ATLAS Note PHYS-98-131(1998); http://www.hep.ucl.ac.uk/atlas/atlfast/
\bibitem{Geant}S. Agostinelli et al., (Geant4 Collaboration), Nucl. Instrum. Meth.
A \textbf{506,} \emph{250} (2003).
\bibitem{R-statistics book}L. Lyons, {}``Data Analysis for Physical Science Students'', Cambridge
University Press, 1991.
\end{thebibliography}
\end{document}